\documentclass{iaus}
\usepackage{graphicx}
\usepackage{natbib}

\title[ADAFs, jets and the SEDs of LINERs] 
{Advection-dominated accretion, jets and the spectral energy distribution of LINERs}

\author[R. S. Nemmen, T. Storchi-Bergmann, M. Eracleous \& F. Yuan]   
{Rodrigo S. Nemmen$^1$, Thaisa Storchi-Bergmann$^1$, Michael Eracleous$^2$ \and Feng Yuan$^3$}

\affiliation{$^1$Instituto de F\'isica, Universidade Federal do Rio Grande do Sul, Campus do Vale, Porto Alegre, RS, Brazil \\ email: {\tt rodrigo.nemmen@ufrgs.br} \\[\affilskip]
$^2$Department of Astronomy and Astrophysics, Pennsylvania State University, 525 Davey Lab, University Park, PA 16802 \\[\affilskip]
$^3$Shanghai Astronomical Observatory, Chinese Academy of Sciences, 80 Nandan Road, Shanghai 200030, China
}

\pubyear{2009}
\volume{267}  
\pagerange{119--125}
\jname{Co-Evolution of Central Black Holes and Galaxies}
\editors{B.M.\ Peterson, R.S.\ Somerville, \& T.\ Storchi-Bergmann, eds.}
\begin{document}

\maketitle

\begin{abstract}
Low-luminosity active galactic nuclei (LLAGNs) represent the bulk of the AGN population in the present-day universe and they trace low-level accreting supermassive black holes. The observational properties of LLAGNs suggest that their central engines are intrinsically different from those of more luminous AGNs. It has been suggested that accretion in LLAGNs occurs via an advection-dominated accretion flow (ADAF) associated with strong jets. In order to probe the accretion physics in LLAGNs as a class, we model the multiwavelength spectral energy distributions (SEDs) of 24 LINERs (taken from a recent compilation by Eracleous et al.) with a coupled accretion-jet model. The accretion flow is modeled as an inner ADAF outside of which there is a truncated standard thin disk. These SEDs include radio, near-IR to near-UV HST data, and Chandra X-ray data. We find that the radio emission is severely underpredicted by ADAF models but can be explained by the relativistic jet. The origin of the X-ray radiation in most sources can be explained by three distinct scenarios: the X-rays can be dominated by emission from the ADAF, the jet, or from both components contributing at similar levels. From the model fits, we estimate important parameters of the central engine of LINERs, such as the mass accretion rate -- relevant for studies of the feeding of AGNs -- and the mass-loss rate in the jet and the jet power -- relevant for studies of the kinetic feedback from jets.
\keywords{black hole physics, galaxies: jets, galaxies: active, accretion, accretion disks}
\end{abstract}

\firstsection 
\section{Introduction}

Thanks in large part to the Palomar survey (e.g., \citealt{ho95, ho97}), we know that most active galactic nuclei (AGN) in the present-day universe have low luminosities, being thus called low-luminosity AGNs (LLAGNs). The bulk of the LLAGN population ($\approx 2/3$) are low-ionization nuclear emission-line regions (LINERs), which are extremely sub-Eddington systems with an average Eddington ratio of $L_{\rm bol}/L_{\rm Edd} \sim 10^{-5}$ \citep{ho09,erac10}. 

The observational properties of LINERs -- and LLAGNs in general -- are quite different from those of more luminous AGNs. 
Regarding the SEDs, LLAGNs seem not to have the big blue bump feature (e.g., \citealt{nemmen06, ho08, yuan09}; but see \citealt{maoz07}) which is one of the signatures of the presence of an optically thick, geometrically thin accretion disk. 
Regarding the emission-lines, LLAGNs typically have weak and narrow Fe K$\alpha$ emission \citep{terashima02} and a handful of LINERs display broad double-peaked H$\alpha$ lines (e.g., \citealt{sb03}); these properties of the emission-line spectrum are consistent with the absence of a thin accretion disk, or a thin accretion disk whose inner radius is truncated at $\gtrsim 100 GM/c^2$. 
Last but not least, with the typical fuel supply of hot diffuse gas (via Bondi accretion) and cold dense gas (via stellar mass loss) available in nearby galaxies, LLAGNs would be expected to produce much higher luminosities than observed on the assumption of standard thin disks with a $10\%$ radiative efficiency \citep{ho09}.
Taken together, this set of observational properties favors the scenario in which the accretion flow in LLAGNs/LINERs is advection-dominated or radiatively inefficient.

Advection-dominated accretion flows (ADAFs; for a recent review see \citealt{nar08}) are very hot, geometrically thick, optically thin flows which are typified by low radiative efficiencies ($L \ll 0.1 \dot{M} c^2$) and occur at low accretion rates ($\dot{M} \lesssim 0.01 \dot{M}_{\rm Edd}$). 
Supermassive black holes are thought to spend $>95 \%$ of their lifes in the ADAF state \citep{hopkins06}, the best studied case being Sgr A* (e.g., \citealt{yuan07}). 

ADAFs are relevant to the understanding of AGN feedback since they are quite efficient at producing powerful outflows and jets, as suggested by theoretical studies (including analytical theory and numerical simulations; e.g., \citealt{nemmen07, nar08}), as well as different observational studies of LLAGNs (e.g., \citealt{ho02, heinz07}). 
In fact, the so-called ``radio mode'' of AGN feedback invoked in semi-analytic and hydrodynamic simulations of galaxy formation (e.g., \citealt{bower06}) would correspond to the ADAF accretion state actively producing jets as explicitly incorporated in some works \citep{sijacki07, oka08}. An alternative and perhaps  more appropriate expression for the ``radio'' feedback mode would be the ``LINER/ADAF'' mode.

It is clear that advances in the understanding of the physical nature of LINERs and LLAGNs are required in order to understand the nature of black hole accretion and feedback in the local universe. The goal of this work is therefore to probe the physics of accretion and ejection in the LINER population, by modeling their nuclear multiwavelength SEDs which provide constraints to physical models for the emission of the accretion flow and the jet.

\section{Data set}

Our data set consists of 24 SEDs of LINERs which include radio (VLA), near-IR -- optical -- UV (HST) and X-ray (Chandra) data with high spatial resolution. These SEDs were selected from a sample of 35 SEDs compiled by \citet{erac10} using two selection criteria: (i) there should be estimates of the black hole mass for the corresponding galaxies and (ii) there should be good X-ray estimates of the photon index and X-ray luminosity. 

Based on these criteria, we can separate the SEDs in two groups: group A, comprising 10 LINERs with the most complete sampling of the SEDs, and group B, for which there is lack of data in some parts of the SED. For illustratiom we list the LINERs in the group A: NGC 1097, M81, NGC 3998, NGC 4143, NGC 4278, M84, M87, NGC 4579, NGC 4594 and NGC 4736.

\section{Models for the continuum emission}

In order to model the LINER SEDs and constrain the properties of their central engines, we adopt the physical scenario which is favored to explain the observational properties of LLAGNs (\citealt{yuan07, ho08}; see Figure 13 of \citealt{ho08} for a cartoon). In this model, the accretion/ejection flow consists of three components: (i) the inner parts of the flow are \textit{advection-dominated}, geometrically thick; (ii) the outer parts of the accretion flow are in the form of a \textit{standard thin disk truncated at a certain transition radius}; (iii) near the innermost parts of the ADAF a \textit{relativistic jet} is launched. More details can be found in \citealt{yuan09} and Nemmen et al. 2010 (\textit{in preparation}).

The radiative processes operating in the ADAF are synchrotron emission, bremsstrahlung and inverse Compton scattering. The truncated thin disk spectrum is simply thermal. The jet contributes with synchrotron emission. Some of the main parameters of the models will be discussed in the section below.

\section{Results}

We describe the results of the spectral fits for only two LINERs of our data set for the sake of brevity: NGC 4374 (M84) and NGC 4594 (Sombrero). These two examples nevertheless are illustrative of our general results. In the SED plots below, the arrows in the IR correspond to upper limits due to significant contamination of the nuclear emission by the host galaxy. The error bars in the near-IR to UV denote the uncertainty due to the range of possible extinction corrections. 

When fitting the SEDs, we explored the parameter space of the accretion/jet models within the range of values plausibly allowed by theory. We avoided to fix the values of some dynamical and microphysical parameters for which there is a substantial uncertainty on theoretical grounds. One example is the parameter $\delta$, which controls the amount of energy dissipated via turbulence that is directly deposited on electrons, for which there is a considerable uncertainty with plausible values in the range $\sim 0.001 - 0.5$ (e.g., \citealt{sharma07}). In all models we considered that only some fraction of the mass accretion rate available at the outer boundary of the accretion flow ends up being accreted, due to mass-loss via winds produced in the ADAF \citep{nar08}.

\begin{figure*}
\centering
\begin{minipage}[b]{0.48\linewidth}
\includegraphics[width=\linewidth]{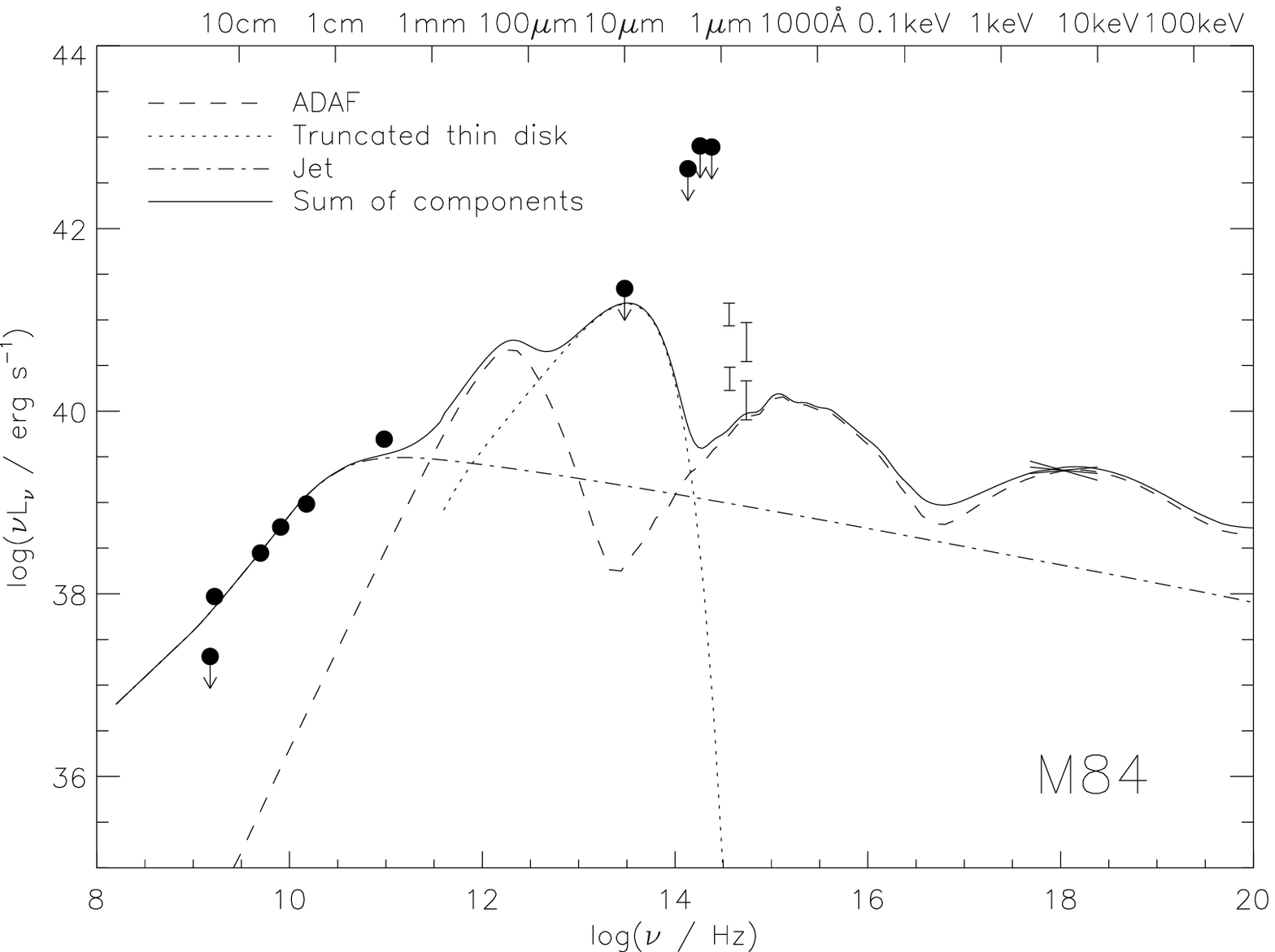}
\end{minipage}
\hfill
\begin{minipage}[b]{0.48\linewidth}
\includegraphics[width=\linewidth]{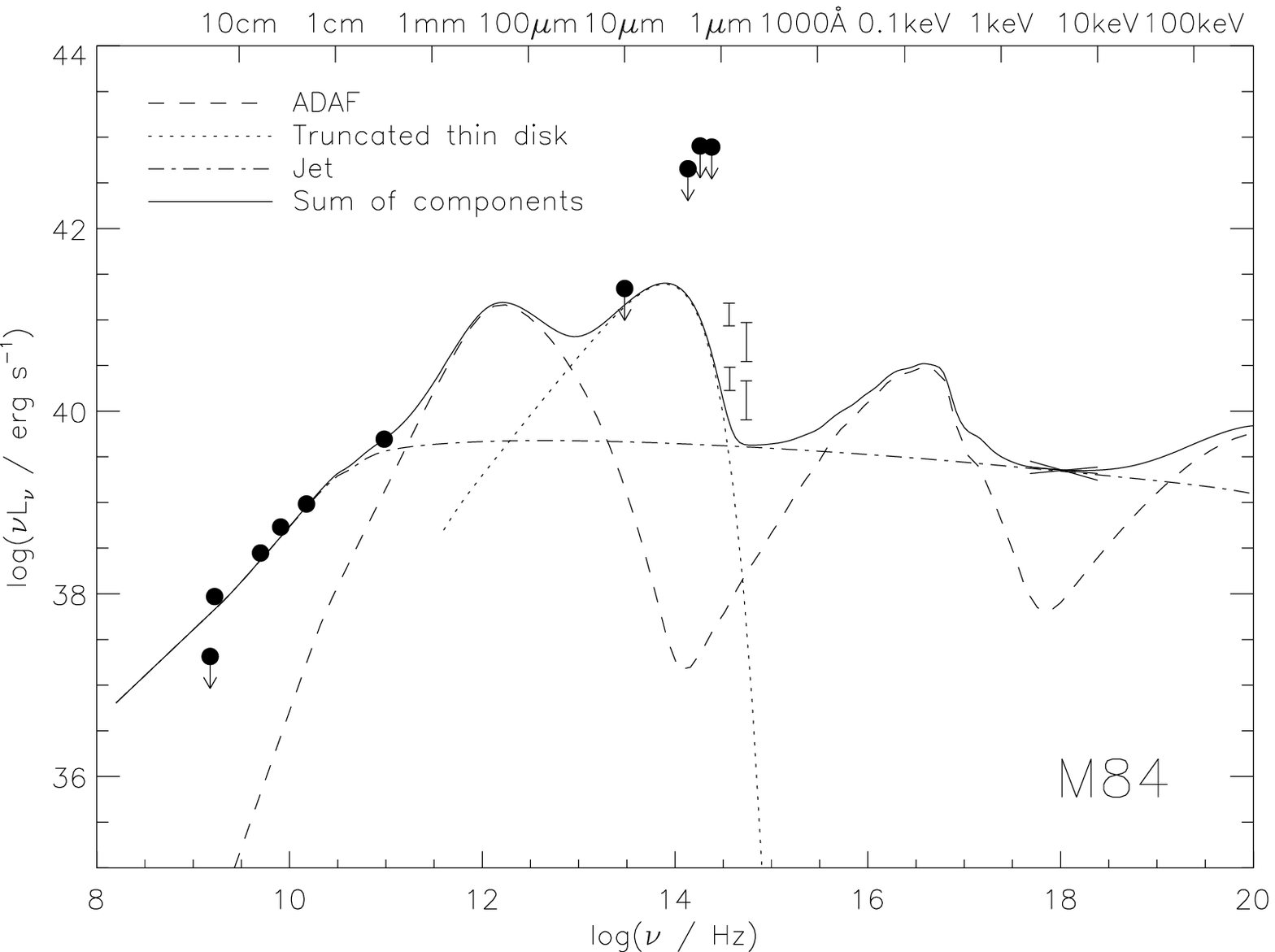}
\end{minipage}
\caption{Models for the SED of M84, including the contribution of the ADAF, truncated thin disk and relativistic jet. Left panel: model in which the ADAF dominates the observed X-ray emission. Right panel: model in which the jet dominates the X-ray output.}
\label{m84}
\end{figure*}

The SED of M84 is plotted in Figure \ref{m84}, together with two different spectral fits. The left panel of this figure shows a model in which the ADAF dominates the emission from the near-IR upwards, particularly the X-ray emission. The thin disk dominates the flux output around $\sim 10 \mu$m and the jet is responsible for the bulk of the core radio emission. For this model, the mass accretion rate supplied at the outer radius $R_{\rm tr}$ of the ADAF is $\dot{M}(R_{\rm tr})=4 \times 10^{-4} \dot{M}_{\rm Edd}$, compatible with the Bondi accretion rate estimated by \citet{pelle05}; furthermore, $R_{\rm tr}=150 R_S$ and $\delta=0.01$. The mass-loss rate in the jet was estimated as $\dot{M}_{\rm jet}=10^{-3} \dot{M}(R_{\rm tr}$).

The right panel of Figure \ref{m84} shows a model for the SED of M84 in which the jet dominates the radio and X-ray emission. Notice that the shape of the X-ray spectrum from the ADAF is not consistent with the data. For this model, $\dot{M}(R_{\rm tr})=1.5 \times 10^{-4} \dot{M}_{\rm Edd}$ (again compatible with the Bondi accretion rate), $R_{\rm tr}=30 R_S$, $\delta=0.3$ and $\dot{M}_{\rm jet}=0.01 \dot{M}(R_{\rm tr})$. 

\begin{figure*}
\centering
\begin{minipage}[b]{0.48\linewidth}
\includegraphics[width=\linewidth]{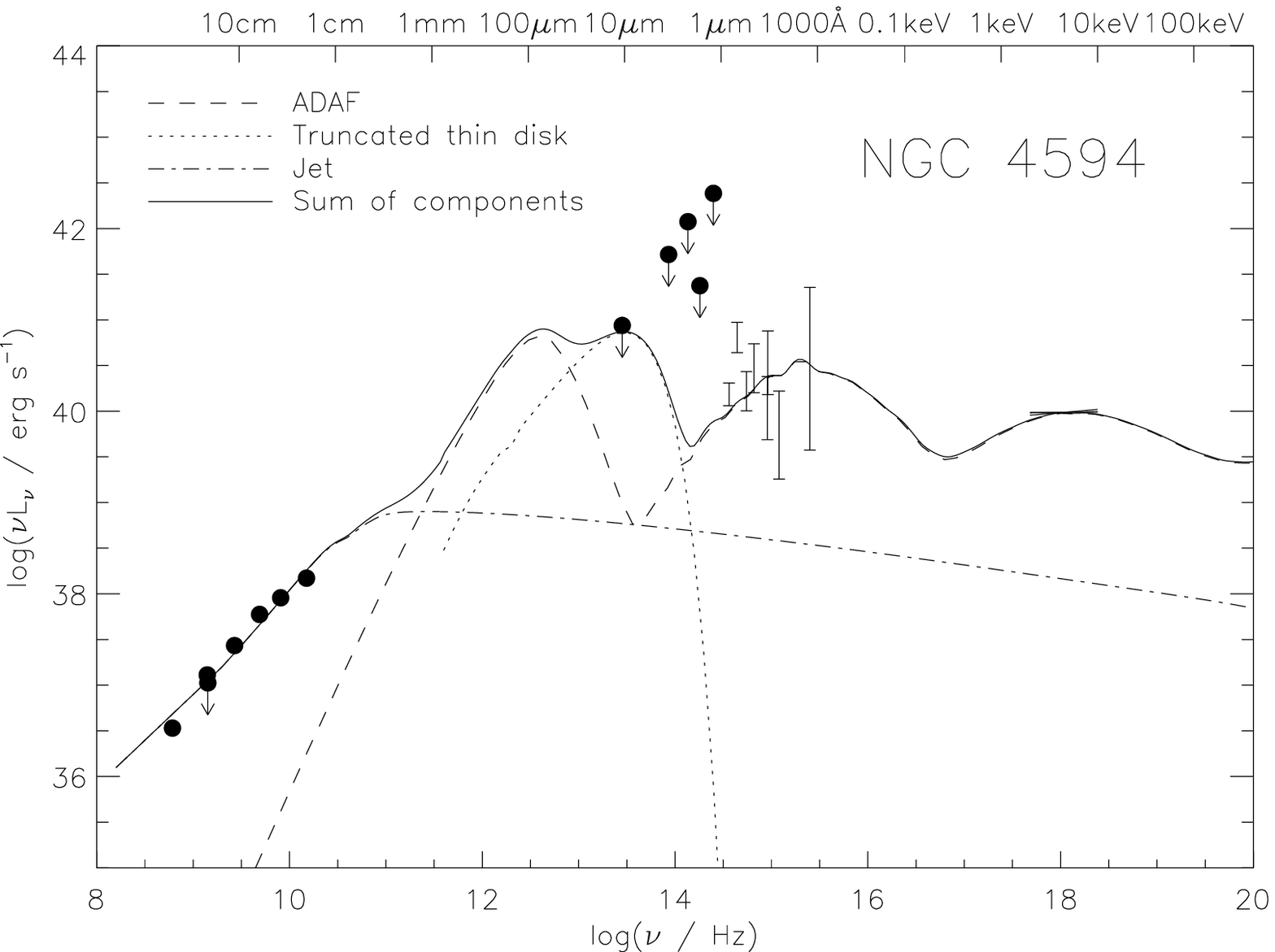}
\end{minipage}
\hfill
\begin{minipage}[b]{0.48\linewidth}
\includegraphics[width=\linewidth]{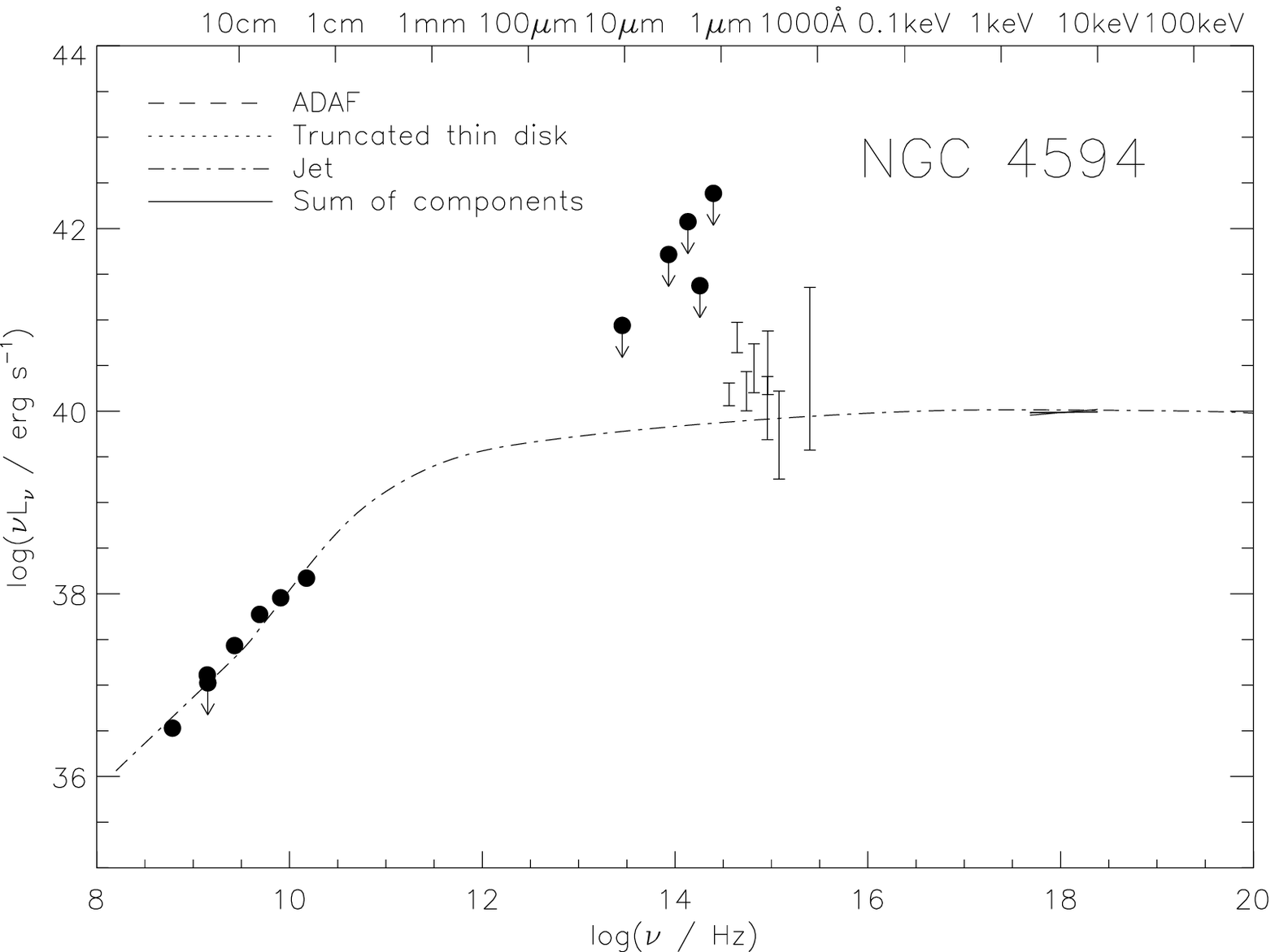}
\end{minipage}
\caption{Same as Figure \ref{m84} for NGC 4594 (Sombrero). Only a jet model is shown in the right panel, for which the mass-loss rate is $\dot{M}_{\rm jet}=4.5 \times 10^{-7} \dot{M}_{\rm Edd}$.}
\label{sombrero}
\end{figure*}

Therefore, we demonstrated that there are two possible types of models which can accomodate the observed SED of M84. In the first type, the emission from the ADAF dominates the observed X-rays; in the second type of model, the jet emission dominates the X-rays. By construction, a third type of model is also possible in which the jet and the ADAF contribute with similar intensities to the high energy emission. 
These results apply also to the other LINERs in our sample. To illustrate this, we also show spectral fits to the nuclear SED of the Sombrero galaxy in Fig. \ref{sombrero}. The left panel shows an ``ADAF dominated X-rays'' model with $\dot{M}(R_{\rm tr})=9.1 \times 10^{-4} \dot{M}_{\rm Edd}$, $R_{\rm tr}=300 R_S$, $\delta=0.01$ and $\dot{M}_{\rm jet}=10^{-3} \dot{M}(R_{\rm tr})$. By varying the microphysical parameters of the jet model, we are also able to obtain a ``jet-dominated X-rays'' model which also successfully explains the entire SED (right panel).

\begin{figure}[t]
\begin{center}
 \includegraphics{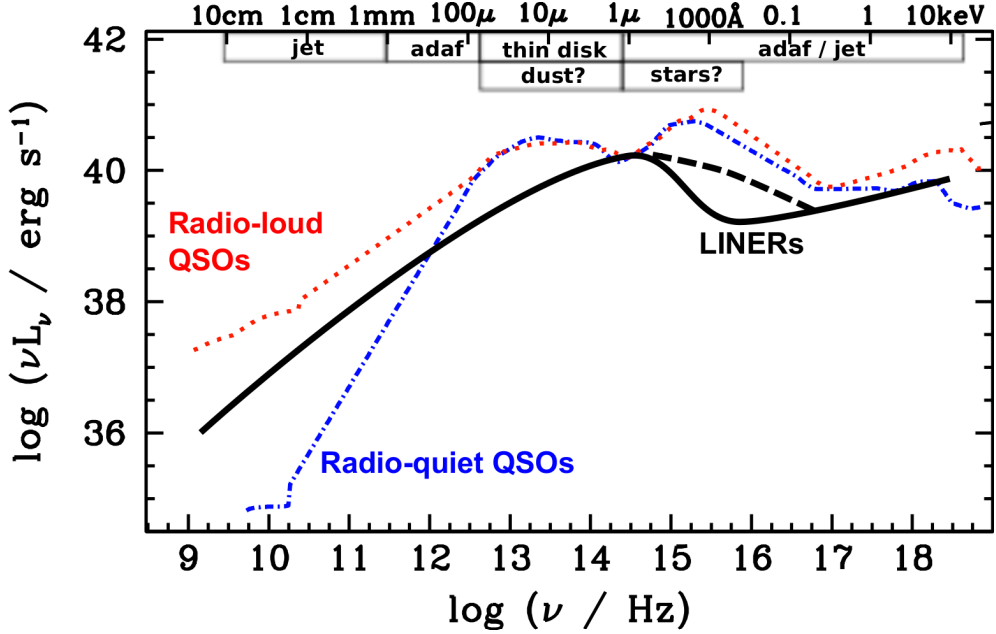} 
 \caption{Qualitative average SED of LINERs (\textit{solid line}) based on \citet{erac10}. The dashed line illustrates the potential effect of extinction corrections. The top part of the plot shows which components of the flow dominate the emission in each waveband. Also displayed are the average SEDs of radio-loud (\textit{dotted line}) and radio-quiet (\textit{dot-dashed line}) quasars from \citet{elvis94} for the sake of comparison.}
   \label{median}
\end{center}
\end{figure}

Figure \ref{median} shows the qualitative average SED of LINERs based on the one computed by \citet{erac10}. From our modelling of the SEDs of LINERs, we are able to unveil the physical nature of the LINER continuum emission in each waveband, as outlined in the upper part of Fig. \ref{median}. The uncertainty about the origin of X-rays in LINERs has been debated in the context of LLAGNs and also Sgr A* \citep{merloni03, falcke04, yuan09}. 
In our case, the root of the uncertainty in the origin of X-rays lies in the uncertainties regarding the microphysics of the hot plasma in the ADAF and the jet (mainly the uncertainty on the value of $\delta$ and the effect of shocks in the jet), which has a major impact in the fitting of the observed X-ray spectrum.
We suggest that monitoring campaigns of the variability of radio and X-ray emission in LINERs, and the comparison of such observations with predictions of jet/ADAF models would help to pin down the nature of the X-ray emission in LLAGNs.

From our modelling of the LINER SEDs with the coupled jet-ADAF model, we are able to constrain important parameters that characterize their central engines:
\begin{itemize}
\item The typical accretion rate available at the outerskirts of the accretion flow is $\dot{M} \sim 10^{-3 \pm 1} \dot{M}_{\rm Edd}$.
\item Typical jet mass-loss rates are in the range $\dot{M}_{\rm jet} \sim 10^{-8} - 10^{-4} \dot{M}_{\rm Edd}$.
\item The values of $\dot{M}_{\rm jet}$ above together with the typical Lorentz factors result in the jet powers $P_{\rm jet} \sim 10^{40} - 10^{43} \ {\rm erg \ s}^{-1}$.
\item Given the bolometric luminosities $L_{\rm bol}$, we have $\left \langle P_{\rm jet} / L_{\rm bol} \right \rangle \sim 10$.
\end{itemize}
Taking the values above, we can roughly estimate the fraction of accreted mass that is channeled into the jets in LINERS, $\left \langle \dot{M}_{\rm jet} / \dot{M} \right \rangle \sim 10^{-3}$. Similarly, we can estimate the efficiency of jet production as $\left \langle P_{\rm jet} / (\dot{M} c^2) \right \rangle \sim 1 \%$, in rough agreement with other estimates for LLAGNs \citep{allen06}.
These values provide useful indicators of the relevant feeding and feedback properties of LINERs, and by extension of the whole LLAGN population; this is particularly relevant in the light of this symposium.

Finally, our work allows us to draw a link between the supermassive black holes in LINERs and the quiescent beast in our galaxy, Sgr A*. By modelling the SED of Sgr A* with current ADAF models, \citet{yuan03} estimated $\dot{M} \sim 10^{-5} \dot{M}_{\rm Edd}$. This value is two orders of magnitude below the typical accretion rate of LINERs that we estimated. Therefore, we could say that if Sgr A* accreted at a rate 100 times its present rate, it would ``light up'' and presumably become a LINER. 

\section{Conclusions}

We were able to successfully model the SEDs of 24 LINERs in the context of a coupled ADAF-jet scenario. While the the radio emission is dominated by the relativistic jet, the ADAF dominates in the band 1 mm -- 100 $\mu$m and the X-ray radiation can be dominated by either the ADAF, the jet or a combination of both. 

We find that strong jets are implied by our modelling, for which the kinetic power considerably exceeds the radiated power ($\left \langle P_{\rm jet} / L_{\rm bol} \right \rangle \sim 10$). Furthermore, we obtained estimates of the fundamental parameters of the central engines of LLAGNs which can be useful in studies of the feeding/feedback properties of AGNs, such as the mass accretion rates, jet powers and mass-loss rates in the jet.

Finally, we would like to point out that our SED models provide a library of templates for the compact emission of LINERs, which can be useful for studies of e.g. the nuclear emission of stellar populations, dust and PAH features. A detailed description of the results above and the SED models will be available in Nemmen et al. (2010), \textit{in preparation}.

\end{document}